\documentclass{PoS}

\title{Spectral modulation of non-Galactic plane Gamma-ray pulsars due to photon-ALPs mixing in Galactic magnetic field.}

\ShortTitle{Spectral modulation of non-Galactic plane pulsars.}

\author{{Jhilik Majumdar$ ^{1}$, Francesca Calore$ ^{2}$, Dieter Horns$ ^{1}$}\\

$ ^{1}$Institute for Experimentalphysics, University of Hamburg.\\
  $ ^{2}$Laboratoire d'Annecy-le-Vieux de Physique $Th\acute{e}orique$,CNRS. \\
  E-mail: \email{Jhilik.majumdar@desy.de}}

\abstract{ Axion like particles (ALPs) are fundamental pseudo scalar particles with properties similar to Axions that have been invoked to solve the strong CP problem in Quantum Chromodynamics. ALPs can oscillate into photons and vice versa in the presence of an external magnetic field. This oscillation of Photon and ALPs could have important implications for astronomical observations, i.e. a characteristic energy dependent attenuation in Gamma ray spectra for astrophysical sources. Here we have revisited the opportunity to search Photon-ALPs coupling in the disappearance channel. We use nine years of Fermi Pass 8 data of a selection of non-Galactic plane Gamma-ray source candidates and study the modulation in the spectra in accordance with Photon-ALPs mixing and estimate best fit values of the parameters i.e. Photon-ALPs coupling constant $( g_{\alpha\gamma\gamma} )$ and ALPs mass$ ( m_{\alpha} )$. For the magnetic field we assume large scale galactic magnetic field models based on Faraday rotation measurements. We find consistent evidence with our prevoius analysis of objects located in the Galactic plane for a modulation of the spectra investigated at the level of 3.66 standard deviations . In the framework of ALPS/photon oscillation, the resulting parameters strongly depend upon the chosen magnetic field model, but seem to be within reach of new experiments (e.g. IAXO).}

\FullConference{7th Fermi Symposium 2017\\
		15-20 October 2017\\
		Garmisch-Partenkirchen, Germany}

\begin{document}
\begin{small}

\section{Introduction}
The probable solution of strong CP problem in Quantum Chromodynamics has been proposed by Pecci \& Quinn postulating a global U(1) symmetry that is spontaneously broken at large energy scale~\cite{Peccei}. Later on Weinberg and Wilczek showed that from this spontaneously broken PQ symmetry, a new pseudo Nambu-Goldstone boson comes into existance, which is also named as axion~\cite{Weinberg}. As this pseudo Nambu-Goldstone boson have a small mass due to non-perturbative effects or explicit symmetry breaking, they could be nice cold dark matter candidates~\cite{Barranco}. Because of low-mass, axions have extremely low decay rates and they hardly make weak interactions with baryonic matter. On the other hand, Axion like particles(ALPs) with even smaller masses may exist and could be detected indirectly by astrophysical observations~\cite{Gorbunov}. ALPs have a nice property that they can oscillate into photons or vice-versa in the presence of magnetic fields~\cite{De Angelis}. Photon-axion conversion induced by intergalactic magnetic fields causes an apparent attenuation in the photon flux of distant sources, depending on the distance of the source, the energy considered and also depending on the tranversal magnetic field along the line of sight~\cite{De Angelis}. A generic feature of axion models is the CP-conserving two-photon coupling, so that the axion-photon interaction is: 
\begin{equation}
        \mathcal{L} \supset - \frac{1}{4}  g_{\alpha\gamma\gamma} F_{\mu\nu} \tilde{F}^{\mu\nu} a = g_{\alpha\gamma\gamma} \vec{E}.\vec{B}a,
\end{equation}

where $a$ is the axion-like field with mass $m_{a}, F_{\mu\nu} $ is the electromagnetic field-strength tensor and $\tilde{F}^{\mu\nu}$ is its dual field , $g_{\alpha\gamma\gamma}$ is the ALPs-photon coupling. Photons, while travelling across the external magnetic field, oscillate with the ALPs state. If the condition, $g_{\alpha\gamma\gamma} Bd \ll 1 $ holds true, the probability of the conversion at a distance $d$ is~\cite{Mirizzi}:

\begin{equation}
        P_{\gamma \rightarrow a}= \frac{g_{\alpha\gamma\gamma}^{2}}{8} \left(| \int^d_0 dz'e^{2\pi i z'/l_{0}} B_{x}(x,y,z') |^{2}+          
|\int^d_0 dz'e^{2\pi i z'/l_{0}} B_{y}(x,y,z') |^{2} \right)\ ,
\end{equation}

Where, $ g_{\alpha\gamma\gamma} $ has the dimension of $(Energy)^-1$ and the parameters are in natural unit. Here, $l_{0}$  is the oscillation length of Photon-ALPs propagation. We are considering the sources with a distance in kpc range, Galactic magnetic field in $\mu G$ ; hence, the photon energy range we take from 100 MeV to 300 GeV.

\section{Galactic Magnetic Field}
Best constraints for large scale Galactic Magnetic Field (GMF) are Faraday rotation measures, polarised synchrotron radiation and also the polarized dust emission from spinning elliptical dust grains. In our analysis we have taken into consideration one of the most comprehensive GMF model: Jansson $\&$ Farrar model (2012)~\cite{Jansson}. This  model consists of three components: Disk component, Halo component and Out of plane component.The disk component is partially based on the structure of the NE2001 thermal electron density model~\cite{Sun}. The disk field is constrained to the x-y plane and defined for Galactocentric radius from 3 kpc to 20 kpc with a 'molecular ring' and eight logarithmic spiral regions. The halo field has a purely toroidal component and separate field amplitudes in the north and south. The out of the plane component, formerly known as X component, is to be asymmetric and poloidal. The components of this large scale magnetic field has been updated with the polarized synchrotron and dust emission data measured with the Planck satellite~\cite{Planck}.

\begin{figure}
\setlength{\unitlength}{.8cm}
\begin{center}
\begin{picture}(15,6)
 \put(-2.5,0){\includegraphics[width=5.5cm]{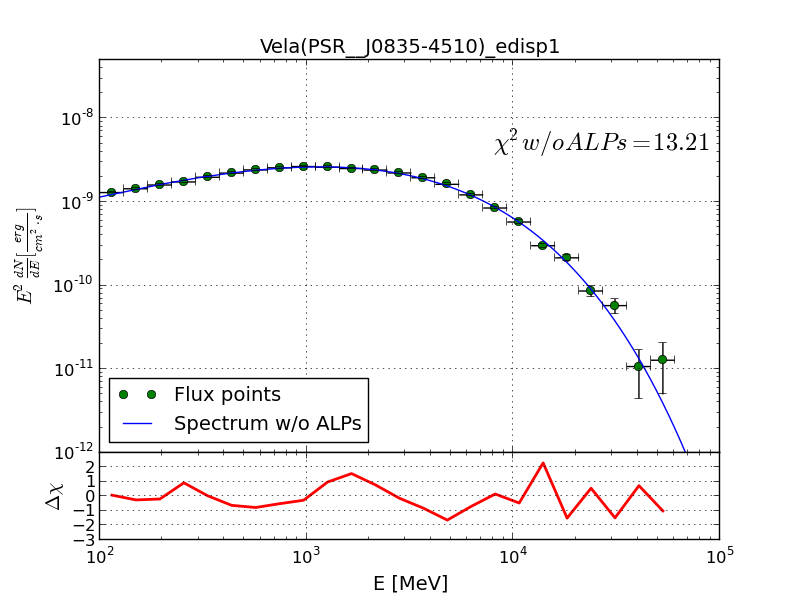}}
 \put(3.8,0){\includegraphics[width=5.5cm]{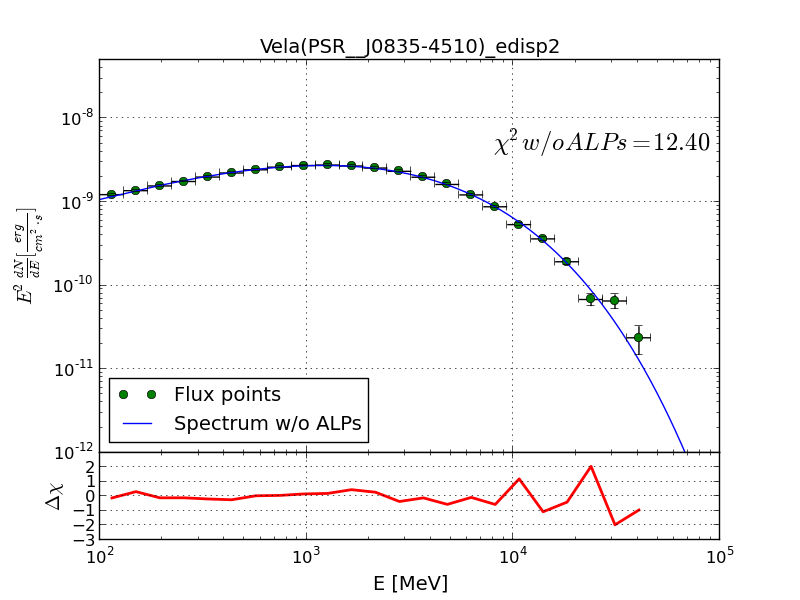}}
 \put(10.2,0){\includegraphics[width=5.5cm]{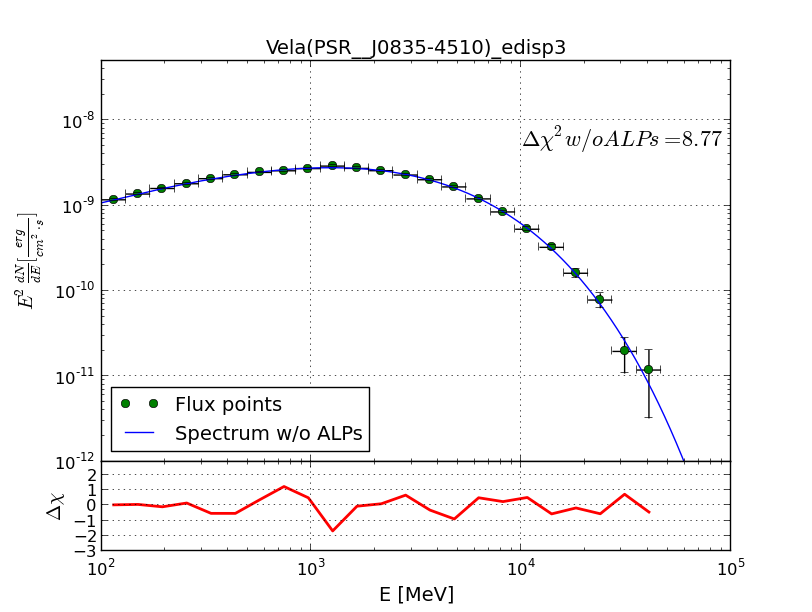}}
 \end{picture}
\end{center}
\caption{A fit of the phase averaged Vela spectrum with a power-law with exponential cut-off model. The $\chi^2$ fit to the differential flux points produces acceptable values under the assumption of a fractional systematic flux error of 2.4\% over the entire energy range, separately shown for EDISP1, EDISP2, EDISP3.}
\end{figure}

\section{Source selection}
In the present work we use gamma-ray data from the Fermi-LAT \cite{Fermi}.We have chosen a list of  twelve bright gamma-ray  pulsars with galactic latitude more than 10 degrees,  listed in table 1.  In order to estimate systematic uncertainties on the observed spectrum we use as expected the Vela pulsar~\cite{Ackermann}. This pulsar is very close; the spectrum is very well measured and does not show any spectral distortion. We use a similar technique that has been done by the Fermi Collaboration to derive the systematics using Fermi-LAT Pass 7 data~\cite{Ackermann}.

\begin {table}
\caption{Gamma-ray pulsar selection list.}
\begin{center}
\begin{tabular}{ |c|c|c|c|c|c| } 
 \hline
 Pulsar name & \multicolumn{1}{|p{3cm}|}{\centering Flux 0.1-300 GeV \\ ( in photons $ cm^{-2} s^{-1}$)} & \multicolumn{1}{|p{1.6 cm}|}{\centering Flux Error \\ $\pm$ } &  \multicolumn{1}{|p{1.6 cm}|}{\centering spectral index} & \multicolumn{1}{|p{2.5cm}|}{\centering Cut-off Energy \\ ( in MeV)} & \multicolumn{1}{|p{1.5 cm}|}{\centering distance \\ ( in kpc)} \\ 
 \hline
 J0007+7302 & 6.48e-08 & 5.72e-10 & 1.20 & 5298.87 & not known \\
 J0030+0451 & 9.82e-09 & 2.75e-10 & 1.28 & 1739.96 & 0.36 \\ 
% J0357+3206 & cell8 & cell9 & cell6 & cell6 \\
 J0614-3329 & 1.74e-08 & 3.57e-10 & 1.36 & 4462.26 & 0.62 \\
 J1231-1411 & 1.77e-08 & 3.78e-10 & 1.18 & 2286.05 & 0.42 \\
 J1311-3430 & 7.54e-09 & 2.47e-10 & 1.90 & 4922.37 & 2.43 \\
 J1836+5925 & 9.93e-08 & 6.53e-09 & 0.88 & 2011.65 & not known \\
 J2055+2539 & 9.03e-09 & 2.78e-10 & 1.04 & 1175.53 & not known \\
 J2124-3358 & 7.47e-09 & 2.58e-10 & 0.89 & 1677.98 & 0.41 \\
 J2214+3000 & 5.64e-09 & 2.17e-10 & 1.20 & 1667.98 & 0.60 \\
 J2229-0833 & 2.94e-08 & 4.56e-10 & 1.80 & 2396.16 & not known \\
 J2241-5237 & 5.21e-09 & 2.03e-10 & 1.25 & 2910.19 & not known \\
 J2302+4443 & 6.53e-09 & 2.34e-10 & 1.19 & 2762.43 & not known \\

 \hline
 
\end{tabular}

\end{center}
\end {table}

\section{Analysis}
 
We use eight and half  years of Fermi-LAT Pass 8 data with P8R2 SOURCE V6 IRFs of twelve bright pulsar candidates, i.e. PSR J0007+7302 et. al and Vela. Pass 8 data has  an  improved  angular  resolution,  a
broader energy range, larger effective area, as well as reduced  uncertainties  in  the  instrumental  response  functions ~\cite{Ackermann}. For spectral modelling of Fermi-LAT sources Enrico binned likelihood optimization technique\cite{Enrico} is performed for 25 energy bins.  
All of the pulsar spectrum is modelled by a power law with  exponential cutoff:
  \begin{equation}
 \frac{dN}{dE}= N_{0} \left(\frac{E}{E_{0}}\right)^{-\Gamma} \exp\left(-\frac{E}{E_{\rm cut}}\right)
 \end{equation}

For Vela we use a power law with sub exponential cutoff :
 \begin{equation}
 \frac{dN}{dE}= N_{0} \left(\frac{E}{E_{0}}\right)^{-\Gamma_{1}} \exp\left[\left(-\frac{E}{E_{\rm cut}}\right)^{\Gamma_{2}}\right]
 \end{equation}

We perform a fit to the data, minimising the $\chi^{2}$ function which provides a measure of goodness of fit~\cite{Ackermann1}~\cite{Jogler} .
We have checked that the log(likelihood) has a parabolic pattern and thus a  $\chi^{2}$ analysis is appropriate. We derive the energy dispersion matrix ($D_{kk_{p}}$) for one energy dispersion event type (EDISP) via the transformation of the number of counts in true energy of a particular energy bin to the number of counts in that bin of reconstructed energy (see Fig.~1) and we fully take it into account in the fit. We investigate the signature of photon-ALPs oscillations, including the effect of oscillations in the predicted spectra:
  
\begin{equation}
\left(\frac{dN}{dE}\right)_{\rm fit} =D_{kk_{p}} .\left(1- P_{\gamma \rightarrow a}\left( E,g_{a\gamma\gamma},m_{a},d\right)\right).\left(\frac{dN}{dE}\right)
\end{equation}

\paragraph{Systematic uncertainties of Vela:} For P8R2\_SOURCE\_V6 event class, systematic uncertainties in effective collection area are derived to be about 2.4 \% for EDISP1, EDISP2 ,  EDISP3 event types for Vela considering the energy range from 100 MeV to 300 GeV(see Fig.~3). We follow the same technique as Fermi collaboration \cite{Ackermann} and do in such a way so that the $\chi^2$ per degrees of freedom(dof) we get $\sim$ 1 and it's an acceptable fit. Systemaics has been imposed in chanel to energy for global fitting and it effects to reduce the $\chi^2$ values for individual pulsars.

\begin{figure}
\setlength{\unitlength}{.9cm}
\begin{center}
\begin{picture}(12,6.5)
 \put(-2.2,0){\includegraphics[width=8cm]{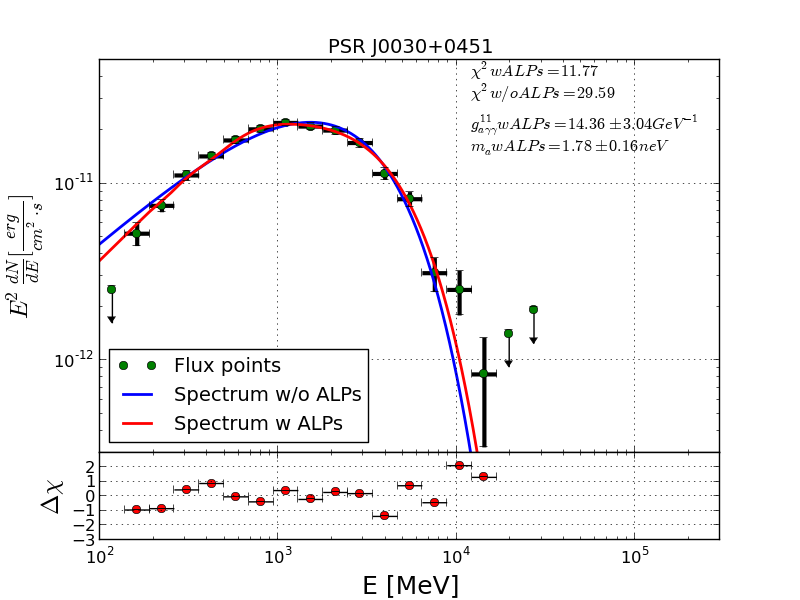}}
 \put(6,0){\includegraphics[width=8 cm]{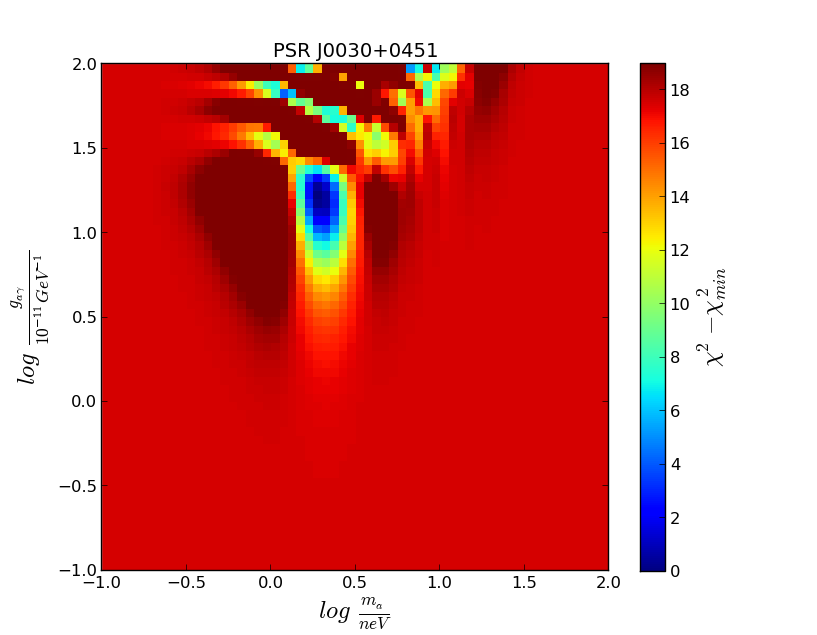}}
 \end{picture}
\end{center}
\caption{Left panel: Best-fit model of the spectrum of PSR J0030+0451 3651. Right panel: The $ \chi^{2}$ scan as function of photon-ALPs coupling and ALPs mass. }

\end{figure}

\begin{figure}
\setlength{\unitlength}{.9cm}
\begin{center}
\begin{picture}(10,6)
 \put(2,0){\includegraphics[width=8cm]{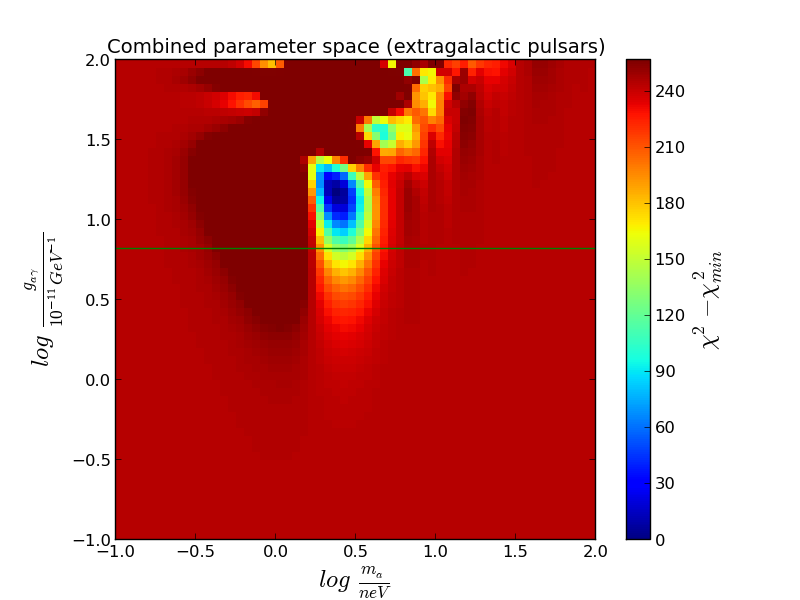}}
  \end{picture}
\end{center}
\caption{Combined $ \chi^{2}$ scan as function of photon-ALPs coupling and ALPs mass.}

\end{figure} 

\section{Result}
Apparent suppression in energy spectrum has been observed in case of Pulsar spectrum due to photon-ALPs mixing in the galactic magnetic field. The spectral feature in disappearance channel generally depends on the distance to the source and transversal magnetic field along the line of sight. We have seen the spectral modulation of the gamma-ray pulsar candidates and it would be explaied by photon-ALPs scenario. In case of  
PSR J0030+0451, the photon-ALPs coupling constant ands ALPs mass we get respectively $14.36 \times 10^{-11} \rm GeV^{-1} $ and 1.78 neV (see Fig.~2). The photon-ALPs mixing phenomena makes a notable improvement in the $\chi^2$  value of the fitted spectrumreducing $\chi^2$ value by 17 from non-ALPs mixing to ALPs mixing case. We can say the fit results   with ALPs parameters is a good fit as the $\chi^2$ per dof is $\sim$ 1 for all the pulsar spectrum. In the fits we also include systematic uncertainties as derived from the Vela pulsar analysis as well as the energy dispersion. Fig.~2(right panel) depicts the blue stripes that corresponds to the photon-ALPs oscillation in tranversal magnetic field. The significance of our result for PSR J0030+0451 has been estimated by F-test 3.66$\sigma$. We make a combined parameter space for all the 12 pulsar candidates (see Fig.~3)  and derive the best parameter space for photon-ALPs coupling constant and ALPs mass. Photon-ALPs coupling constant ($g_{\alpha\gamma\gamma}$) we get $14.12 \times 10^{-11} \rm GeV^{-1} $ and ALPs mass ($m_{\alpha}$) we calculate 2.23 neV.  The irregularities in pulsar spectrum can be explained by detailed analysis of photon-ALPs mixing in the galactic magnetic field, but we should be very cautious to choose the magnetic field model as the mixing parameters are very much model dependent. 
\end{small}

\end{document}